\documentstyle[11pt,paspconf,epsf]{article}

\begin{document}

\def\ltsima{$\; \buildrel < \over \sim \;$}\
\def\simlt{\lower.5ex\hbox{\ltsima}}

\title{Studying the Evolution of Field Galaxies Using NICMOS/HST Parallel Imaging And Grism Data}
\author{L. Yan, P.J. McCarthy, L.J. Storrie-Lombardi, R.J. Weymann}
\affil{The Carnegie Observatories, 813 Santa Barbara St., Pasadena, CA 91101 Email: lyan@ociw.edu}

\begin{abstract}
We present results from our analysis of F160W NICMOS Parallel images. 
These data cover $\sim$~9~sq. arcminutes and reach 3$\sigma$
depths of H$=$ 24.3 $-$ 25.5 in a $0.6''$ diameter aperture with
integration times of 2,000 to 13,000 seconds. We derive the 
first deep H band galaxy counts. The slope of the counts for H$<$ 20
is 0.31, consistent with various K-band measurements from the Keck 
telescopes. The measured number counts vs. magnitude relation is reasonably
well fitted with no-evolution models with a low $\Omega$ value.
The half-light radii of the galaxies declines steeply with apparent magnitude
and reaches the NIC3 resoltion limit at H$=$23.5. Deep ground-based
VRI imaging of one NICMOS field has revealed an extremely red galaxy with
R$-$H $=$ 6 and H of 18.8. Our analyses of the grism data show 
that we can reach
3$\sigma$ flux limits of of $1\times 10^{-16}$ to $2\times10^{-17}$
ergs/sec/cm$^2$ for integration times of 2,000 to 21,000~seconds. 
We have detected a total of 33 emission line galaxies. 
The comoving number density is $\rm \sim 2\times 10^{-4}
Mpc^{-3}$. The detected emission lines are probably H$_\alpha$~6563\AA. 
Thus, the derived star formation rates, without extinction correction, 
are $10 - 163 M\odot$ per year
for galaxies at redshifts between 0.7 and 1.9.

\end{abstract}


\section{Introduction}
The NICMOS parallel observations, taken in parallel with one of the other
science instruments on HST, has provided us for the first time
a wealth of data at near  infrared wavelengths with HST resolution.
Small background at wavelengths of 0.8$\mu$ and 1.6$\mu$ and HST
high angular resolution make NICMOS a very efficient instrument 
in studying the faint galaxy population at high redshifts. 

The NICMOS parallel imaging and grism observations were both made
with Camera 3 with a field of view $\sim$ 52$''\times$ 52$''$.
The imaging data were taken with broad band filters F110W and F160W
at 1.1$\mu$ (J band) and 1.6$\mu$ (H band).
The grism data has a spectral resolution of 200 per pixel and 
covers wavelength regions from 0.8$\mu$ to 1.2$\mu$ (G096)
and 1.1$\mu$ to 1.9$\mu$ (G141). Currently, for high 
galactic fields (l $>$ 20 degrees), the NIC3 parallel 
imaging data cover an area of 
$\sim$167 sq. arcminutes and the grism data $\sim$65 sq. arcminutes.
The final expected total areal coverage is close to
200 sq. arcminutes for the imaging data and 100 sq. arcminutes for
the grism data. 

Despite the less than optimal focus, we achieve depths that are very
close to the normal NIC3 performance. With exposure times of 2,000 to 13,000
seconds we reach 3$\sigma$ of $\rm H = 24.3 - 25.5$ in an aperture of 
0.6$''$ diameter in the direct imaging data. The measured FWHM for stellar
objects is $\sim$ 0.23$''$ in the imaging data. The grism data reaches 
3$\sigma$ flux limits of 1$\times 10^{-16}$ to 2$\times 10^{-17}$ ergs/sec
/cm$^2$ for integration times of 2,000 to 21,000 seconds.

\section{Deep H-band Galaxy Counts and Half-light Radii}

The first result from these 
deep images is the number magnitude relation (Yan et al. 1998). 
The deviations in faint galaxy number counts 
from the no-evolutionary predictions 
mostly reflect the dynamical and luminosity evolution, although
number counts also probe the geometry of the Universe.
The relative importance of the two primary forms
of evolution, density and luminosity evolution, 
can only be properly assessed with spectroscopic redshifts.
The near-IR pass-bands, however, are better suited than visible colors
to purely photometric surveys as they are less sensitive to
star formation and extinction. The weak dependence of
the K-correction on Hubble type and its slow change
with redshift further enhance
the value of observing at wavelengths beyond $\sim 1\mu$m.

\subsection{Observations and Data Reduction}
We have completed analyses of a small set of H band images, which 
covers $\sim$ 9 sq. arcminutes at high galactic latitutes ($>$ 25 degrees).
Our data reduction approach relies heavily on McLeod's NICREDv1.5
package (McLeod 1998). We made observed sky+dark frames fro each read in 
the multiaccum sequence by computing the median of all of the independent
fields. These median sky+dark
images were subtracted from each read and each pixel was then 
corrected for nonlinearities and cosmic ray events. The individual 
linearized and cleaned images were then corrected for the flat field response,
block replicated to a 2x2 finer scale, shifted by integer ($0.''1$) pixel
shifts and combined with a final 3$\sigma$ cosmic ray rejection applied. The 
deepest of our NIC3 parallel images reaches a 1$\sigma$ surface brightness
limit of 26 H magnitude per square arcsecond. The 3$\sigma$ point-source
detection limits for our fields range from H of 24.3 to 25.5 in a 
$0.''6$ diameter aperture. The 50\%\ completeness limits are H of 23.5 to 24.5.
More detailed information can be found in Yan et al. 1998.
Figure 1 shows an image with a typical integration time from
the parallel program. It has 4500~seconds and reaches a 1$\sigma$
surface brightness limit of 25.7 magnitude/sq. arcsec in H band.
The deepest field to date, shown in Yan et al. 1998, has
an additonal 0.5 magnitudes of depth.  

\begin{figure}[h]
\centering \leavevmode
\epsfysize = 7truecm \epsfbox{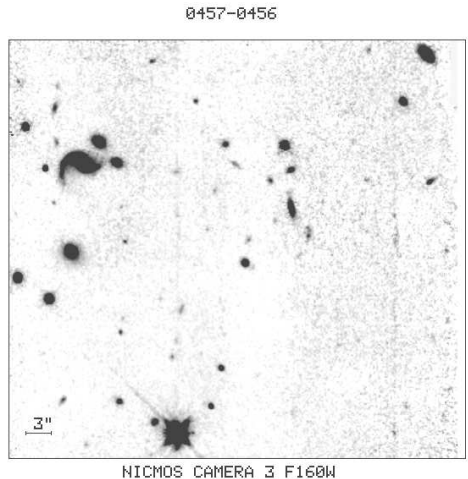}
\caption{An image in F160W filter with a typical integration time from the NICMOS
Camera 3 Parallel program}
\label{fig-1}
\end{figure}

\subsection{Galaxy Photometry and Incompleteness Modeling}
We performed the object detection and
photometry using SExtractor version 1.2b10b
(Bertin \&\ Arnouts 1996). 
We used isophotal magnitudes to define total magnitudes for galaxies
with isophotal diameters $> 0.6^{''}$. 
For faint galaxies with isophotal diameters $< 0.6^{''}$, we use their 
aperture magnitudes. These two magnitudes are both corrected to
a 2$^{''}$ diameter aperture to get the total magnitudes. 
The aperture corrections are obtained from 
an average of 15$-$30 galaxies in the corresponding magnitude range.

The raw counts have been corrected for false detections and incompleteness.
We selected several well detected galaxies
from an image, dimmed them by various factors, and added these images into
the original image at {\em random locations}. We then apply the same
detection and photometry algorithms as in the original
analyses. Detailed information can be found in Yan et al. (1998). We adopted a 50\%\
completeness as the cutoff and the simulations show that this corresponds
to depths of H$=$ 23.8 to 24.8 for our chosen fields. As described
in Yan et al. (1998), the false detection rate is of order 5\%.

\subsection{Results and Discussion}

In Figure 2, we plot
our raw and corrected counts with the open circles and solid dots
(Yan et al. 1998).
We have converted K band magnitudes to the NICMOS F160W magnitude
using the models from Gardner (1998). 
The open diamond and triangle represent the K band galaxy counts
by Bershady et al. (1998) and Djorgovski et al. 1995 respectively.
The solid and dashed lines correspond to the no-evolution models by
Gronwall \&\ Koo (1995) with $q_0$ of 0.5 and 0.05. This figure
shows that our number counts are consistent with what measured
in the deep K-band images from the Keck telescope. 
Our data cover 9 sq. arcminutes, whereas the two
Keck K band surveys only imaged roughly 1$-$1.5 sq. arcminutes.

\begin{figure}[h]
\plotfiddle{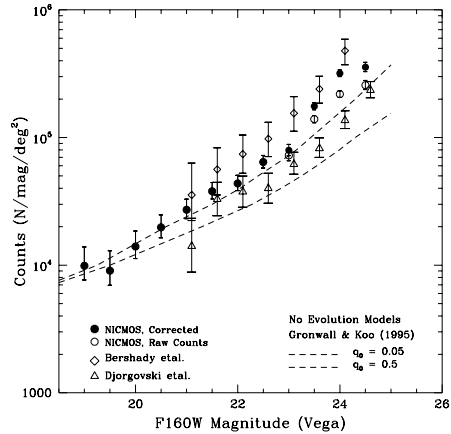}{4.5truecm}{0.0}{30}{30}{-120}{-50}
\caption{The galaxy number counts vs. magnitude
relation down to H magnitude of 24.5 (50\%\ incompleteness limit).
The open and filled circles are the raw and corrected counts,
respectively.}
\label{fig-2}
\end{figure}

The slope of
our counts is 0.31$\pm$ 0.02 for
for 20$<$H$<$24.5. We find no significant change in the slope for H $< 24.5$.
At an H magnitude of 24.5, the galaxy number-magnitude relation does not
strongly departure from no-evolution models, in contrast to what we 
see in the optical bands. 
The integrated number of galaxies H$\leq 24.5$, including the incompleteness
corrections, is $4 \times 10^5$ per sq. degree, or 2$\times 10^{10}$ galaxies
over the entire sky. This is about 3 times larger than
the total implied from integration of the local luminosity function (Lin et al. 1996) to
0.01L$^\ast$ over an all-sky co-moving
volume for ($\Omega_0$, $\Omega_\Lambda$) $= (1, 0)$ cosmological model.
If the bulk of the galaxies that we detect at
H $\sim 24$ are faint dwarf galaxies with
luminosity much less than $\rm L^\ast$
at low redshifts instead of $\rm L^\ast$ galaxies at $z > 3$,
the counts slope of $0.31$
implies a H band luminosity function $\Phi(L) \propto L^{-1.78}$.

\begin{figure}[h]
\plotfiddle{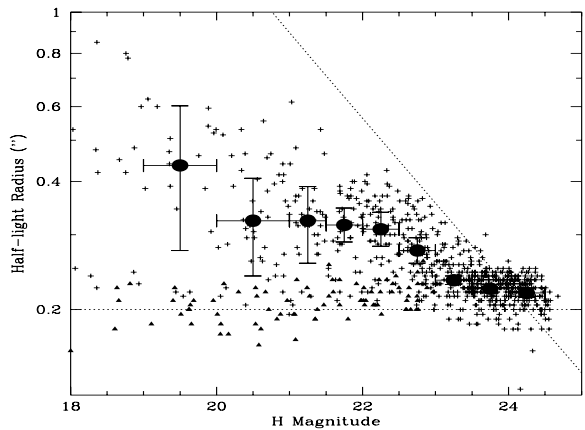}{5truecm}{0.0}{40}{30}{-150}{-55}
\caption{The half-light radii vs. H magnitudes for all of the detected galaxies.
See the text for the detailed explanation.}
\label{figure-3}
\end{figure}

We charcterize the size of the detected objects in terms of their
half-light radii. Figure 3 plots the derived half-light radii versus 
H magnitude for all detected galaxies. The solid symbols are the
median sizes in bins of 0.5 or 1 magnitude.
The dotted line corresponds to a uniform surface brightness of
22 mag/arcsec$^2$. The star symbols represent the measurements from a 
globular cluster field. At H of 23.5 we reach the HST resolution limit.
For H $<$ 22, fainter galaxies have smaller apparent sizes.

\section{Galaxy Colors and Extremely Red Objects}

Figure 4 is J$-$H vs. H diagram. The open circles
are the data measured from the NIC3 images; the solid line indicates the
color-magnitude track for no-evolution elliptical galaxies; the dashed line is
for star forming galaxies with a flat spectral energy distribution (SED).
The solid dots are the median colors. In this plot , the object indicated
by an arrow has a V$-$H $>$ 6, with V$>$ 25, R $=$ 24.8, I $=$ 23.2, 
F110W(J) $=$ 20.5 and F160W(H)$=$ 18.8.
In the H-band image, this object is well resolved and appears fairly
regular. Its morphology is well fitted by a $r^{1/4}$ law profile of an elliptical galaxy.
This is in contrast to the distorted morphology reported for 
another ERO by Graham \&\ Dey (1996).  

\begin{figure}[h]
\plotfiddle{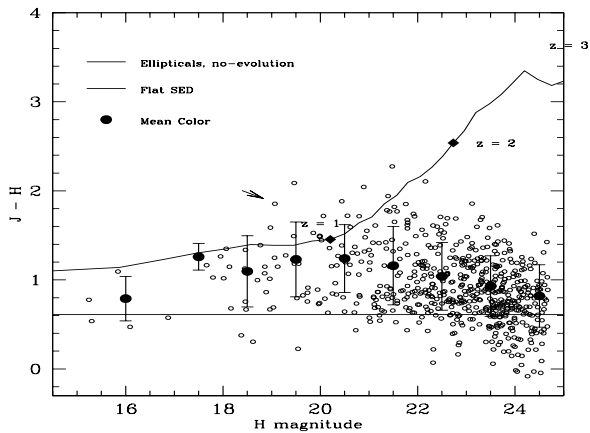}{5truecm}{0.0}{40}{30}{-150}{-57}
\caption{The J$-$H color versus H magnitude relation. The object marked by the
arrow is an ERO with R$-$H $=$ 6.}
\label{figure-4}
\end{figure}

Although several groups have discovered
a number of EROs with R$-$K $>$ 6 (McCarthy, Persson \&\ West 1992; 
Hu \&\ Ridgeway 1994), the statistics of EROs is still very poor and
the nature of these objects remains unclear. These objects could be 
old ellipticals formed in the monolithic collapse at high redshifts (Hu \&\ Ridgway 1994);
they could also be dust shrouded star forming galaxies (Graham \&\ Dey 1996; 
Cimatti et al. 1998). Combining ground-based optical photometry with NICMOS
parallel observations, our survey will search for EROs over a large area of sky 
($\sim$ 200~sq. arcminutes.) with an unprecedented depth.


\section{Preliminary Results from Grism Data}

Currently there are $\sim$65 sq. arcminutes of high galactic latitute 
fields which have grism observations covering $\lambda$ = $1.1\mu - 1.9\mu$.
We have reduced all of the data and the detailed analyses are under way.
The grism data reaches 3$\sigma$ flux limits of $1\times 10^{-16}$ to $2\times10^{-17}$
ergs/sec/cm$^2$ for integration times of 2,000 to 21,000~seconds. We have detected
a total of 33 emission line galaxies with line fluxes of $2\times 10^{-15}$ ergs/s/cm$^2$ to
$2\times10^{-17}$ ergs/s/cm$^2$. These detected lines are likely H$_\alpha$~6563\AA, which
put these emission line galaxies in the redshift range between 0.7 and 1.9.
The derived H$_\alpha$ luminosity ranges from $2\times 10^{41} - 2\times 10^{43}$ ergs/s.
The implied star formation rates, without extinction correction, are $10 - 163 M\odot$ per year
with the median of 24$\rm M\odot/yr$.
Figure 5 shows the spectrum in both 2D and 1D form for a typical emission line galaxy. 
This galaxy has H magnitude of 20.5.

\begin{figure}[h]
\plotfiddle{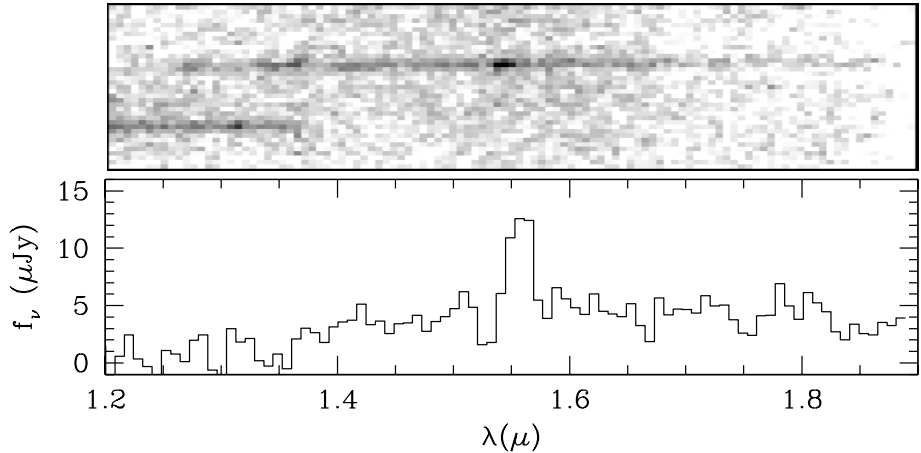}{5truecm}{0.0}{70}{70}{-220}{-80}
\caption{An example of emission line galaxy. The galaxy is at z $\sim$ 1.37 for 
the line being H$_\alpha$. This galaxy has H magnitude of 20.5.}
\label{figure-6}
\end{figure}

\acknowledgments
We thank the NICMOS team at the Space Telescope Institute, and 
acknowledge useful discussions with B. McLeod, I. Smail,
J. Gardner, H. Teplitz, R. Thompson, M. Rieke and David Hogg. 
This research is supported by grants from Space Telescope Science Institute, AR-07972.01-96A,
GO-7499.01-96A and PO423101.

\end{document}